# Plasmonic nanoprobes for stimulated emission depletion microscopy

Emiliano Cortés[1*], Paloma A. Huidobro[1*], Hugo G. Sinclair[2‡], Stina Guldbrand[2], William J. Peveler[3], Timothy Davies[4], Simona Parrinello[4], Frederik Görlitz[2], Chris Dunsby[2,5], Mark A. A. Neil[2], Yonatan Sivan[6], Ivan P. Parkin[3], Paul M. French[2], Stefan A. Maier[1]

1- Experimental Solid State Group, Department of Physics, Imperial College London, SW7 2AZ, UK
2- Photonics Group, Department of Physics, Imperial College London, SW7 2AZ, UK
3- Department of Chemistry, 20 Gordon Street, University College London, WC1H 0AJ, UK
4- MRC Clinical Sciences Centre (CSC), Imperial College London, Du Cane Road, London, W12 0NN, UK
5- Centre for Pathology, Imperial College London, Du Cane Road, London, W12 0NN, UK
6- Unit of Electro-optics Engineering, Ben-Gurion University, Beer-Sheba 8410501, Israel

‡ Present address: Institute of Structural and Molecular Biology, Darwin Building, Gower Street, University College London, WC1E 6BT, UK

*Corresponding authors: E.C. e.cortes@imperial.ac.uk and P.A.H p.arroyo-huidobro@imperial.ac.uk

## Abstract

Plasmonic nanoparticles influence the absorption and emission processes of nearby emitters due to local enhancements of the illuminating radiation and the photonic density of states. Here, we use the plasmon resonance of metal nanoparticles in order to enhance the stimulated depletion of excited molecules for super-resolved microscopy. We demonstrate stimulated emission depletion (STED) microscopy with gold nanorods with a long axis of only 26 nm and a width of 8 nm that provide an enhancement of the resolution compared to fluorescent-only probes without plasmonic components irradiated with the same depletion power. These novel nanoparticle-assisted STED probes represent a ~2x10$^3$ reduction in probe volume compared to previously used nanoparticles and we demonstrate their application to the first plasmon-assisted STED cellular imaging. We also discuss their current limitations.

# Introduction

The diffraction limit has ceased to be a practical limit to resolution in far-field microscopy, following the demonstration of STED [1,2,3], RESOLFT [4] and localisation microscopies [5,6,7] and the subsequent development of a plethora of super-resolved microscopy techniques [8]. In particular, stimulated emission depletion (STED) nanoscopy, which builds on the advantages of laser scanning confocal microscopy, is a powerful technique for super-resolved imaging in complex biological samples including live organisms [9,10]. STED microscopy uses stimulated emission to turn off the spontaneous fluorescence emission of dye molecules, typically overlapping a focused excitation beam with a "doughnut" shaped beam that de-excites emitters to the ground state everywhere except for the area within the centre of the doughnut, thus providing theoretically diffraction-unlimited resolution in the transverse plane by reducing the full-width half-maximum (FWHM) of the point spread function. By increasing the power of the depletion beam the emission region can be can be drastically reduced - theoretically allowing for sub-nanometre resolution - with resolutions of less than 10 nm being demonstrated [11]. The scaling of resolution with the square root of the depletion beam power means that relatively high-power lasers are typically used for STED nanoscopy. In practice, however, the use of high-power irradiation can result in problems such as photobleaching of the fluorophores and phototoxicity, and so the achievable resolution is compromised by the need to limit the intensity of the depletion laser radiation. Furthermore, high power lasers can add cost and complexity to STED microscopes and so the requirement for high power depletion beams presents challenges for parallelizing STED measurements [12,13] in terms of the lasers required, thus, limiting the potential for faster super-resolved imaging.

To some extent the issue of photobleaching can be addressed with the development of new, more robust fluorescent labels such as quantum dots [14] as labelling agents but they still require relatively intense depletion-beams and present concerns around phototoxicity. In order to reduce the required intensity of the depletion beam, we recently proposed the use of plasmonic nanoparticles (NPs) whose localized surface plasmon resonances (LSPRs) are spectrally tuned to the depletion beam wavelength [15,16,17]. Briefly, materials that present plasmonic resonances enable an intense sub-wavelength concentration of light to below the diffraction limit, mediating electromagnetic (EM) energy transfer either from the far to the near field or vice versa. Thus, these elements can be considered as optical nanoantennas and are key instruments in the conversion of free-space light to nanometre-scale volumes [18,19]. For metal NPs at their LSPR, their scattering cross section can be much larger than their physical cross section, which allows them to efficiently capture light. This light is then concentrated in the vicinity of the nanoparticle and can be used, for example, to deplete excited fluorescent molecules, among others [20,21,22,23]. Thus, if the LSPR of the metal NP is centred at the depletion-laser wavelength, the input depletion-power requirements for STED microscopy can be met at much lower intensities than normally used.

Our recent demonstration of the proof of concept of nanoparticle-assisted STED (NP-STED) microscopy utilized 160 nm core@shell (silica@Au) spherical particles, with a plasmon resonance at ~800 nm and fluorescent dyes embedded in the silica core [17]. However, such large NPs are not practically useful for most studies in biology. In order to facilitate their incorporation into cells and to best represent the underlying labeled structures of interest when using them to read out biological processes, the size of the plasmonic-fluorescent probes should be as small as possible [24]. Typical resolution achieved in STED using fluorescent dyes for bio-imaging is below 50 nm [25], which can be taken as an upper limit for the sizes of experimentally useful plasmonic NP labels. A further issue for the core@shell nanoparticles used in our proof of concept demonstration was the observation of unwanted background light arising from gold luminescence. This limited the resolution achievable and we found it necessary to implement time-gated detection in order to supress this unwanted background light [17].

For NP-STED microscopy implemented with a mode-locked Ti:Sapphire laser to provide the depletion radiation, plasmonic-probes utilising the fluorescence of the most popular near infrared STED dyes should be designed to have a plasmon resonance at ~780 nm and ideally be smaller than 50 nm. Au and Ag are the most common materials employed in plasmonics to achieve near-infrared and visible-wavelength LSPRs, due to their well-behaved optical constants [26]. By tuning the shape and size of the particles it is possible to shift their plasmon resonance over the whole visible and the near-infrared spectrum [27]. However, as Ag and Au spheres of ~50 nm diameter have their plasmon resonances at ~450 and ~550 nm respectively, achieving a LSPR at 780 nm with a 50 nm maximum-size for metallic (Ag or Au) particles requires the use of non-spherical particles. Anisotropic particles, such as gold nanorods (AuNRs), emerge as a suitable shape to fulfil these requirements as their plasmon resonance can be tuned by adjusting their aspect ratio (length/width) [28]. Specifically, AuNRs with an aspect ratio of ~3.25 ensure a longitudinal plasmon resonance at 780 nm in oil; usually used as a refractive index matching media for microscopy measurements. Thus, if the length and width of the AuNRs are reduced while keeping this ratio constant, it is possible to achieve smaller plasmonic-probes suitable for STED microscopy [29].

Here, we demonstrate this extension of NP-STED to smaller, anisotropic particles by synthesizing new plasmonic-probes based on 26x8 nm fluorescently-labelled AuNRs and show that we can achieve a resolution improvement of ~50% using STED microscopy at low depletion intensities (1.5 MW/cm$^2$) for which a control experiment using fluorescent beads without plasmonic enhancement presents a much weaker (<10%) improvement in resolution. These new plasmonic nanoprobes for STED are ~2000 times smaller in volume compared to the nanoparticles we reported previously [17] and we have used them to label adult neural stem cells for STED microscopy at low depletion powers. We also note that the reduction (~1000x) in the amount of metal of these NP means that we no longer detect the unwanted background light from gold luminescence.

## Results and Discussion:

We start by describing the functionalization and characterization of the small plasmonic probes, designed to perform as labelling agents for low-power STED super-resolution imaging. As already mentioned, we synthesized small, fluorophore-coated AuNRs, whose longitudinal plasmon resonance is matched with the depletion-beam wavelength (780 nm in our STED microscope, see Figure S1 for more information about the STED-FLIM set-up). In order to conjugate the fluorescent dye (STAR 635P) onto the surface of the AuNRs we used a biotin-streptavidin system (Figure 1(a)). This linker provides a separation of ~5 nm between the fluorophore and the metal surface of the nanorods in order to avoid fluorescence quenching

[30,31]. It also enables us to incorporate a molecule providing chemically specific binding such as phalloidin, which is used to provide tight and selective binding to actin in cells (for further details on the functionalization see the Methods section). Two centrifugation steps were carried out in order to fully purify the AuNRs. Spectral characterization of the supernatant and pellet confirmed the removal of excess dye.

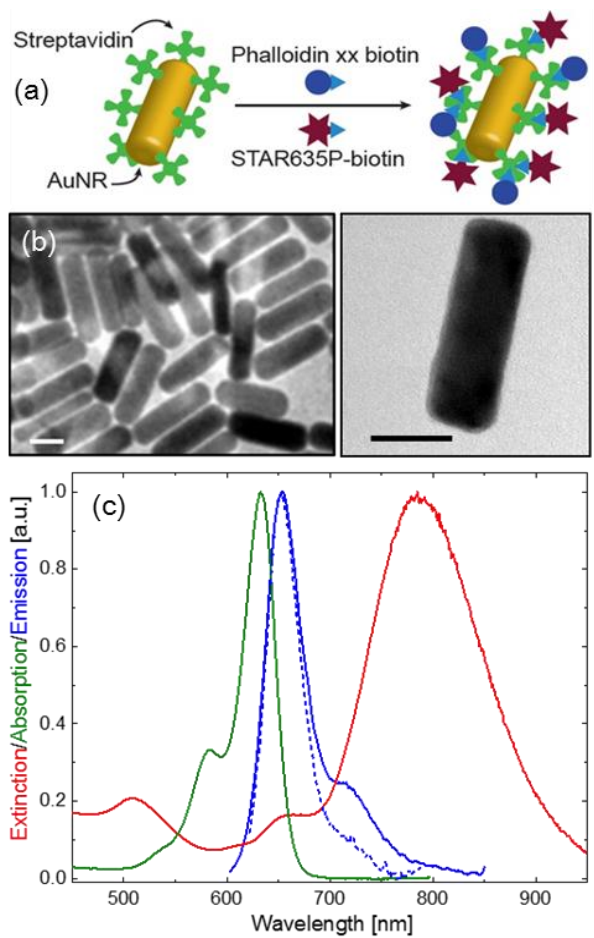

*Figure 1: (a)* Scheme of the functionalization procedure employed to synthesize the plasmonic-fluorescent-probes. Streptavidin coated AuNRs, featuring approximately 15 streptavidin proteins per nanorod, were incubated with biotinylated phalloidin (20 µM in MeOH, 50 µL) and biotinylated STAR 635P dye (0.1 mg/mL in DMSO, 14 µL). After 3 hours the mixture was centrifuged twice, before resuspending the pellet in $H_2O$. *(b)* Representative TEM images of the AuNRs. Scale bars: 10 nm. *(c)* Optical characterization of the AuNRs. Green (blue) line is the absorption (emission) spectrum of the dye (STAR 635P). The blue dashed line shows the emission of the dye once bound to the AuNRs (excited at 600 nm). The red line is the extinction spectrum of a STAR 635P-AuNRs thin film, measured in ProLong Gold media and showing the transversal and longitudinal plasmon resonances at 520 nm and 780 nm, respectively.

The average size of our AuNRs (26±4 nm length and 8±1 nm width) was determined by TEM (N >100 particles - see Figure S2 for distributions). Figure 1(b) shows representative TEM images of the post-functionalized AuNRs. The extinction spectrum of the ~3.25 aspect ratio AuNRs measured in ProLong Gold media shows LSPRs centered at 520 nm (transversal mode) and 780 nm (longitudinal mode) - see red line in Figure 1(c); in agreement with calculations (see Figure S3). We also present in Figure 1(c) the absorption (green line) and emission spectrum (dashed blue line; excited at 600 nm) of the STAR 635P conjugated to the AuNRs. The corresponding emission spectrum of the free STAR 635P dye is also shown (solid blue line). The fluorescence lifetime was found to be the same for the AuNR and the free dye (see Figure S4) and the emission spectra are similar apart from an apparent decrease at the red edge that we attribute to emitted photons being absorbed by the AuNRs at those wavelengths.

In order to characterize the electromagnetic response of the employed AuNRs, their extinction cross section was calculated by means of finite element method simulations (see Methods section for more details). Figure 2(a) presents the calculated extinction cross section of an AuNR (under plane wave illumination), showing the longitudinal LSPR at 780 nm and the transverse LSPR at 520 nm, in close agreement with the measured spectrum (see Figure 1(c)). The inset panel (Figure 2(a) confirms that the longitudinal LSPR supported by the AuNRs at 780 nm corresponds to a dipolar plasmon resonance excited by the electric field component of the incident wave parallel to the particle's long axis. This panel exemplifies the large field enhancement at the LSPR, reaching values over 35 close to the rod's tips surface, where the field enhancement is at its maximum. The spatial region of a few nm away from the metal is known to cause quenching of the dye through non-radiative decay channels in the metal. As discussed above, by linking the dyes to the AuNRs surface with the biotin/streptavidin conjugation, they are kept at an average distance of 5 nm from the metal, thus, avoiding the quenching of the plasmonic-fluorescent nanoprobes [30,31]. At this distance, the still-appreciable field enhancement produced by the AuNR (see Figure 2(b)) causes and enhancement of the depletion intensity experienced by the fluorophores.

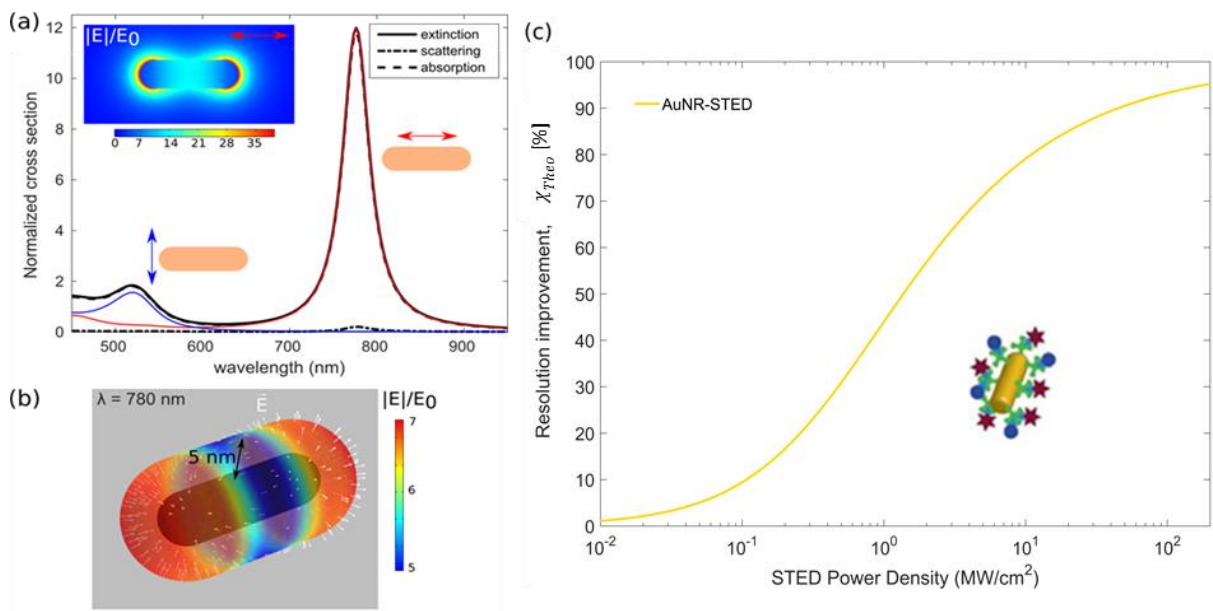

*Figure 2: (a)* Simulated spectrum of AuNRs (26x8 nm) in oil (n=1.47, to reproduce the experimental conditions). The solid lines show the extinction cross section for incident light polarized along the rod's long axis (red), short axis (blue) and the corresponding for circularly polarized light (black). The absorption and scattering contributions to the extinction are given with dashed and dotted-dashed lines, respectively. Inset panel: on-resonance electric field enhancement ($\lambda$ = 780 nm) for under excitation with the **E** component along the rod's axis (note the wave's power corresponds to a circularly polarized plane wave). *(b)* Field enhancement and vector field on a surface 5 nm away from the rod's surface, the average value of the fluorophore distance to the rod. *(c)* Orange line shows the calculated resolution improvement (%) curve $\chi_{Theo}$ as a function of the depletion (STED) average laser power densities using our fluorescent-AuNRs. The fluorescent dye STAR 635P was assumed for the calculations.

To demonstrate the potential of the AuNRs for NP-STED, we functionalized a glass cover-slip to attach the AuNRs to the surface (see Methods for further details) and compared the resolution improvement relative to confocal microscopy for the AuNR and for 20 nm diameter crimson red beads as a function of the power of the depletion-beam. Assuming that the depletion beam pulse length, $\tau_{STED}$, is smaller than the fluorescence lifetime of the fluorophore, $\tau_{S1}$, the resolution achievable with pulsed STED depends on the wavelength, $\lambda$, the numerical aperture (NA) of the objective, the peak intensity of the depletion beam, $I_{STED}$, and the saturation intensity of the fluorophore, $I_{sat}$, as given by [16,32]:

$$FWHM_{STED} = \frac{FWHM_0}{\sqrt{(1 + \emptyset)}}$$

where

$$FWHM_0 = \frac{0.51\,\lambda}{NA}$$

and

$$\emptyset = \frac{I_{STED}\,\tau_{STED}}{I_{Sat}\,\tau_{S1}}$$

Hence, we can write the STED resolution improvement with respect to confocal microscopy as $\Gamma_{res} = \sqrt{1 + \emptyset}$. In the case of NP-STED, we can introduce an extra factor describing the spatially-averaged improvement in resolution due to the nanoparticle. The resolution improvement factor of NP-STED microscopy [16] can, therefore, be written as:

$$\Gamma_{res} = \sqrt{1 + \Gamma_p \emptyset}$$

where

$$\Gamma_p = \frac{\Gamma_I}{\gamma}$$

Here, $\Gamma_I$ is the local increase of the depletion intensity at the fluorophores location due to the electromagnetic field enhancement (at the STED wavelength), and $\gamma$ is the enhancement of the decay rate of the dye, which is modified due to the Purcell effect (at the emission wavelength). An increase in the fluorescent decay rate is detrimental as it reduces the total amount of collected photons. For this reason, the metal NPs were designed to be out of resonance at the dye emission wavelength, such that $\gamma \approx 1$, which was confirmed by the lifetime measurements of the sample (see Figure S4) [33].

The resolution improvement of the NP–STED system is, therefore, estimated by means of the electromagnetic intensity enhancement, $\Gamma_I$. Such enhancement factor can be calculated for the AuNRs under plane wave illumination. Indeed, Foreman *et al.* [34] have shown that such an estimate provides a crude but easy alternative for exact numerical simulations, such as those performed in Ref [15]. On the other hand, since the fluorophores are randomly positioned over the rod at ~5 nm from its surface and with randomly oriented dipole moments, the averaged value of the local enhancement $\bar{\Gamma}_I$, has to be considered. As shown in Figure 2(b), the field enhancement at that distance from the rod ranges from 5 to 7. Averaging the local field intensity over the fluorophore's positions and orientations yields an intensity enhancement factor $\bar{\Gamma}_I \approx 8.5$ (see Methods). Yet, the effective enhancement is reduced because of the lower emission cross section of the hybrid STAR 635P-AuNR emitter, see Fig. 1(c), which gives rise to a higher saturation level. With this, we characterize the resolution improvement factor in percent via the expression [17].

$$\chi_{Theo} = 100 \cdot \left(1 - \frac{1}{\Gamma_{res}}\right)$$

This calculated resolution improvement curve ($\chi_{Theo}$) is shown in Figure 2(c) for the STAR 635P-AuNRs. This can be directly compared to the measured resolution improvement:

$$\chi_{Exp} = 100 \cdot \left(1 - \frac{FWHM_{STED}}{FHMW_{conf}}\right)$$

By comparing the FWHM of the bright spots (i.e. single fluorescent-AuNR or single fluorescence bead) observed from the AuNRs under confocal or STED imaging, it is possible to estimate this resolution improvement (%) factor, $\chi_{Exp}$, experimentally for a given STED depletion-power, as exemplified in Figure 3(b) (see Figures S6 and S7 for more examples). Figure 3(a) shows the experimental resolution improvement, $\chi_{Exp}$, for fluorescent-AuNRs (orange crosses) and for the non-plasmonic probes (20 nm crimson fluorescent beads - green crosses - see experimental methods for details). This was determined for selected AuNR/beads across the sample as a function of the power of the depletion beam; see examples in Figures 3(b) and 3(c).

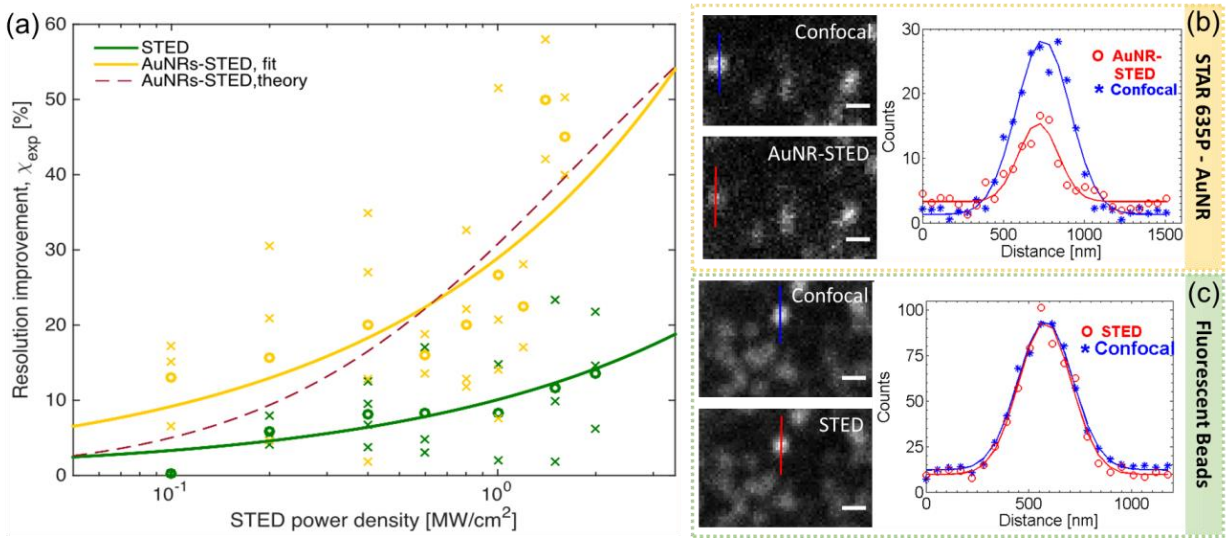

*Figure 3: (a)* Orange/green crosses are experimental measurements for fluorescent-AuNRs/fluorescent beads at different depletion average power densities and orange/green open circles are the average resolution improvement for AuNRs-STED/fluorescent beads-STED. Lines are fits for the average behaviour. The dashed red line shows the resolution improvement based on the calculated field enhancement and emission cross section of the hybrid STAR 635P-AuNR emitter. *(b)* Fluorescent-AuNR confocal (top) and NP-STED (bottom) images and the corresponding fittings showing the resolution improvement for the selected rod (marked with lines in the images) using depletion laser intensities of 0.2 MW/cm$^2$. *(c)* Bottom panel: under the same conditions, using 20 nm fluorescent beads, there is little improvement in resolution for such a low depletion power (0.2 MW/cm$^2$). ProLong Gold is used as index-matching media for AuNRs and TDE (97%) for the fluorescent beads sample. Scale bars in images correspond to 500 nm.

Although there is a significant dispersion in the measurements for both systems (which in the case of AuNRs may arise from clustering of the NPs, from a finite AuNRs size-distribution or from thermally-induced motion of the AuNR as discussed below), there is a clear trend indicated by the average values found for each depletion power (open circles in Figure 3(a)). For depletion-powers as low as 1.5 MW/cm$^2$, we observed a resolution improvement of ~50% for the AuNRs - in very good agreement with the numerical estimates of Figure 2(c) - while for the fluorescent beads this value is below 10% under the same experimental conditions. This implies an improvement of 44% for AuNR-STED compared to dye-based STED at this depletion power. Equivalently, when using the AuNRs, a given resolution improvement is attained by about 10 times lower intensity compared with the control measurement. In fact, since crimson beads have a lower saturation intensity than STAR 635P (see Methods section), a comparison of the AuNRs against beads labelled with STAR635P would be expected to show roughly a 50% larger improvement in resolution.

As well as these new plasmonic probes for STED being ~2000 times smaller in volume than the previously used silica@Au particles [17], the average near field intensity experienced by the fluorophores in the case of the AuNRs is ~2 times larger at the depletion wavelength (see Methods section and Figure S5). The amount of gold in each AuNR is also greatly reduced (1000 times) compared to the Au shell of previously used particles. This reduces the gold luminescence below our detection limit and is not apparent in the emission decay profiles (see SI). Thus we do not need to apply time-gated detection to prevent the spurious luminescence compromising the STED performance, as was necessary for the core@shell nanoparticles [17] and which decreased the detected fluorescence signals. Unfortunately, while the shape and small size of the chosen NPs provides a large electric field enhancement and the possibility to incorporate them into cells (as shown next), this is accompanied by an increase in the absorption of the metal. In practice, this leads to heating and thermally induced movement of the AuNRs that becomes significantly challenging for STED microscopy with depletion powers over ~2 MW/cm$^2$ (see further discussion in the SI). This clearly presents a practical challenge for NP-STED, as further discussed in the conclusions.

Finally, we incorporated fluorescently labelled AuNRs into adult neural stem cells in order to perform super-resolution imaging inside the cell (see Methods for details). Figure 4 shows an initial study, where the resolution improvement for NP-STED can be clearly seen across the image for depletion powers as low as 0.5 MW/cm$^2$. However, it is important to take into account the anisotropic nature of these AuNRs that would be bound to actin and the potential impact on the resolution improvement of out-of-plane orientations. In order to investigate this possibility, we performed calculations on the "effective" point spread function (PSF) of the depletion beam and the expected emission PSF, depending on the 3D orientation of the AuNRs (see Figures S9-S11). Our calculations show that the degree of improvement in resolution can vary depending on the specific orientation of the AuNRs. This last fact can account for the brighter/dimmer spots shown in Figure 4(c) compared to 4(b). Although further work is required to improve the labeling of relevant intracellular structures, this proof of concept experiment nevertheless indicates the potential for lower power STED microscopy of biological samples [22].

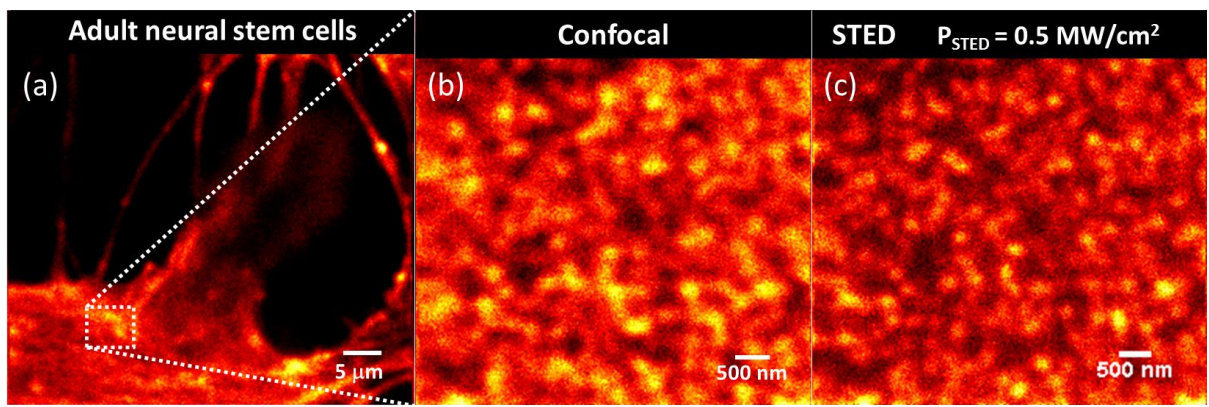

*Figure 4: a)* *AuNRs incorporation into adult neural stem cells.* *b)* *Confocal versus* *c)* *NP-STED images inside the cell (zoom from the marked region in a). Depletion-beam power for the NP-STED case was set at 0.5 MW/cm$^2$.*

## Conclusions:

While many different types of nanoparticles (made out of, e.g. silica, carbon, polymers, quantum dots, etc.) have been utilised as probes for fluorescence imaging [35], super-resolved bioimaging with nanoparticles has only recently been accomplished - using quantum dots [14]. In this report, we have shown the possibility to extend the NP-STED concept to significantly smaller nanometric plasmonic probes that we have shown can be incorporated into cells. Employing the near-field enhancement properties of fluorescently labelled 26x8 nm AuNRs, we achieve a resolution improvement of 50% compared to 10% for STED microscopy of fluorescent beads with the same depletion power. We further note that the smaller size of these AuNRs compared to the core@shell NPs [17] eliminates the issue of unwanted background light from gold luminescence. This demonstrates the potential of plasmon-assisted STED to realize super-resolution at low depletion powers.

Currently, the main challenge for NP-STED for bioimaging is the heating of the NPs and surrounding medium upon illumination, which can lead to thermally-induced motion of the NPs and ultimately their destruction at higher depletion powers [17]. It may be possible to mitigate these issues to some extent by optimizing the duration and repetition rate of the STED pulses and taking advantage of the rapid equilibration of the NPs temperature, which we calculate to be ~6 ns (see figure S8) [36]. For instance, by increasing the duration of the pulses (i.e. 1 ns) while maintaining the total pulse energy, it is possible to reduce the peak temperature rise by over 50% (see Figure S8) – although time-gated detection may then be necessary to optimize the STED microscopy resolution. Another possibility would be to reduce the depletion intensity required to deplete the dye by working at shorter wavelengths [37,38]. Specifically, designing anisotropic NPs with a LSPR centered at ~750 nm, where the cross-section for stimulated emission depletion of STAR635P is approximately 3 times higher than at 780nm, could reduce the intensity required to achieve a given STED resolution by the same factor of ~3, thus, further reducing the heat generated. Engineering smaller plasmonic particles would also reduce the amount of heat generated, however, the field-enhancements obtained with these particles will be lower. This can be compensate by positioning the emitter closer to the metal surface, in which case their quantum yield and fluorescense lifetime will decrease. However, the photon output may be compensated by a reduction of bleaching.

Finally, as a further development of the NP-STED concept, we envision the possibility of combining plasmonic particles with quantum dots (as fluorophores) in order to have a robust probe in terms of photobleaching [14] with the plasmon modes enhancing the depletion of the emission [39,40], thus requiring less power for STED imaging.

## Experimental Methods:

**Synthesis and Functionalization of AuNRs**: Streptavidin coated AuNRs (Nanopartz, 150 μL), featuring approximately 15 streptavidin proteins per nanorod, were incubated with phalloidin-xx- biotin (20 μM in MeOH, 50 μL) and biotinylated STAR 635P (Abberior) dye (0.1 mg/mL in DMSO, 14 μL). After 3 hours the mixture was centrifugally washed at 9660 x *g* twice before resuspending the pellet in 1 mL of H$_2$O (18 MΩ, ultrapure). Calculated molar quantities: Streptavidin – 2.28 x $10^{-10}$ moles; phalloidin-xx-biotin – 1 x $10^{-9}$ moles; STAR 635P – 1 .1 x $10^{-9}$ moles. We estimate a labelling ratio of 7 fluorophores per nanorod.  A control sample was made up in the same way without the dye, and with 150 μL of phalloidin-xx- biotin to ensure correct rejection from the excitation and STED lasers (no fluorophores) and discard interference

(i.e. scattering) from the metallic nanoparticles. Absorption and emission spectra were measured in solution using water as solvent. $I_{sat}$ ~15MW/cm² was estimated for STAR 635P from its lifetime, τ, and extinction coefficient, σ, given by the data sheet [http://www.sigmaaldrich.com/life-science/cell-biology/detection/abberior-dyes.html], as $I_{sat} = k \cdot h \cdot c/(\lambda_{STED}\sigma)$, where $k$ is the spontaneous decay rate, $k$ =1/τ, and σ is the extinction coefficient at the STED wavelength.

**Fixation of AuNRs to the substrate:** AuNRs were fixed on glass substrates by soaking the substrates in PBS 3 times, 10 minutes each. After that, a mixture of 1mg/ml biotinylated BSA and 0.1 mg/ml BSA was left in contact with the substrates for 3 hours followed by 3 washing steps with PBS. AuNRs were diluted 50 times in PBS (final OD close to 1) and then put in contact with substrate for 2 hours followed by washing 3 times with PBS. Finally, biotinylated STAR 635P (1 mM) was used to label the fixed rods in the substrate and then washed two times with PBS.

**Preparation of fluorescent bead sample**: 20 nm diameter fluorescent beads (Crimson beads, Invitrogen, Carlsbad, CA, US) with excitation/emission maxima at 625/645 nm were diluted 1:1000 in MilliQ and applied to Poly-L-lysine coated coverslips. They were mounted in TDE (97%) in order to match the refractive index of 1.52 for the immersion oil. $I_{sat}$ ~2 MW/cm², was estimated for crimson beads from Ref [41].

**Scattering of gold nanoparticles:** The electromagnetic response of the nanoparticles is characterized by the extinction cross section under plane wave incidence. Extinction is given by the sum of scattering and absorption, which were calculated using the finite element method [Comsol Multiphysics, https://www.comsol.com/comsol-multiphysics]. The scattering cross section is obtained from the power scattered by the nanorod and the absorption cross section from the electromagnetic energy absorbed by the particle. The calculation is performed for two different incident polarizations (electric field parallel to the long or short axis of the rod), and circular polarization is then considered. The results shown in Figure 2(a) correspond to cross sections normalized to the illuminated area of the particle. The permittivity of gold, $\varepsilon_{Au}(\omega)$, was taken from the data given in reference [42].

**Plasmonic enhancement factor of STED resolution:** As discussed in the main text, the NP-STED resolution improvement is given by $\Gamma_{res} \approx \sqrt{1 + \Gamma_P \phi}$, where $\Gamma_P \approx \bar{\Gamma}_I$. In order to calculate $\bar{\Gamma}_I$ we perform a series of averages of the dipole orientations of the AuNR. We start from a simulation with a linearly polarized plane wave, since only the electric field component of the circularly polarized beam that lies along the rod's long axis excites the longitudinal LSPR (note that we take the appropriate normalization factor into account and that all the field enhancement values given in this paper correspond to a circularly polarized wave). Next, we average the intensity enhancement at the fluorophore position, $\Gamma_I(r) = |\mu \cdot E_{SP}(r)|^2$, over random dipole orientations ($\mu$ is the unit dipole moment in the direction of the fluorophore's dipole and $E_{SP}(r)$ the near field enhancement at the LSPR). Finally, in order to take into account the inhomogeneous field distribution provided by the AuNRs we average over random positions over the AuNR, $\bar{\Gamma}_I = 1/A \int dA \cdot \Gamma_I(r)$, with $A$ being the area of the surface 5 nm away from the AuNR, as depicted in Fig. 2(b). After performing all the averages, we obtain a plasmonic intensity enhancement factor of 8.5, which corresponds to an average field enhancement for the fluorophores of 3. Note that in case not all the rods lie on the plane, as could potentially be for the bioimaging example shown in Figure 4, the averaged plasmonic intensity enhancement factor would change. This is discussed further in the SI.

**STED microscopy set-up:** The STED-FLIM microscope employed in this study is a development of that described previously [43,44] utilising a single 80 MHz repetition rate mode-locked Ti:Sapphire laser (Spectra-Physics, Mai Tai HP) tuned to 780 nm to provide the depletion beam and to generate a supercontinuum in a microstructured fibre from which the excitation pulses were selected using a bandpass filter (628/40 nm). Further details on the experimental set-up are presented in the SI.

**STED microscopy:** STED images were taken by first acquiring a confocal image by scanning 10 frames and adding them together to give a final image. The acquisition time was 0.41 s for each 64x64 frame, where the pixel size was 55.96 nm. Once the confocal image was acquired, the depletion beam was switched on and a STED image was acquired by scanning and summing 10 frames. The depletion laser pulses were set to arrive 10 ps after the excitation pulses. The average power of the depletion beam was measured before the tube lens of the microscope and the STED pulse intensity ($I_{STED}$) was calculated using $I_{STED} = k \frac{P_{STED}}{A_{STED}}$, where $k$ = 0.3, $A_{STED}$ is the STED focal area and $P_{STED}$ the corresponding peak power [45] obtained by dividing the pulse energy by $\tau_{STED}$.

**FLIM microscopy:** FLIM (Fluorescence Lifetime IMaging) data was acquired in confocal imaging mode by accumulating five 64x64 frames, where the pixel size was 55.96 nm and the acquisition time was 0.41 s for each frame. Fluorescence intensity images were analyzed using Fiji and MatLab 2013a (The Mathworks, Inc, Natick, Massachusetts, US) and FLIM data was analyzed using our in-house open source tool, *FLIMfit* [46].

**Cell stain, fixation and measurements:** Adult neural stem cells were treated for 15 minutes with 4% paraformaldehyde (PFA) in PBS buffer and permeabilised for 10 minutes with 0.5% triton-x100 detergent in PBS at room temperature. One litre of PBS buffer was made out of: 8g NaCl, 0.2g KCl, 0.2g $KH_2PO_4$, 1.15g $Na_2HPO_4$. After that the cells were blocked for 30 minutes with 3% bovine serum albumin (BSA) in PBS at room temperature. Finally, cell's staining was carried out with a dilution of AuNRs-Atto647-phalloidin probes (resonant at 820 nm in ProLong Gold media) in a solution of 3% BSA in PBS (in order to reach an optical density of ~1) for 1 hour at room temperature, washed 3 times for 10 minutes with PBS and mounted in ProLong Gold anti-fade reagent. Depletion wavelength was set to 820 nm in this case.

## Supplementary Information:

STED set-up, Nanorods size-distribution, LSPR dependence on AuNRs aspect ratio and embedding medium, Lifetime measurements, Nanoshells vs. nanorods, Additional NP-STED images, Photothermal effects, AuNRs-STED anisotropy.

# Acknowledgements:

This project was supported by the MRC Next Generation Optical Microscopy Initiative, (MR/K015834/1), EPSRC Reactive Plasmonics project EP/M013812/1, the Royal Society and the Leverhulme Trust. EC and PAH are supported by Marie-Sklodowska Curie fellowships. WP is supported by an EPSRC Doctoral Prize Fellowship (EP/M506448/1). HS acknowledges a CASE Ph.D studentship from the UK Engineering and Physical Sciences Research Council (EPSRC) and the Diamond Trading Company. YS acknowledges the support from the People Programme (Marie Curie Actions) of the European Union's Seventh Framework Programme (FP7/2007-2013) under REA grant agreement no.333790, as well as the support of the Alumni program of the International Newton Fellowship program of the Royal Society and of the Israeli national nanotechnology Initiative. SAM acknowledgers the Lee-Lucas Chair in Physics. We want to acknowledge Dr. Yannick Sonnefraud for initial discussions on this project.

# Supplementary Information

## STED set-up

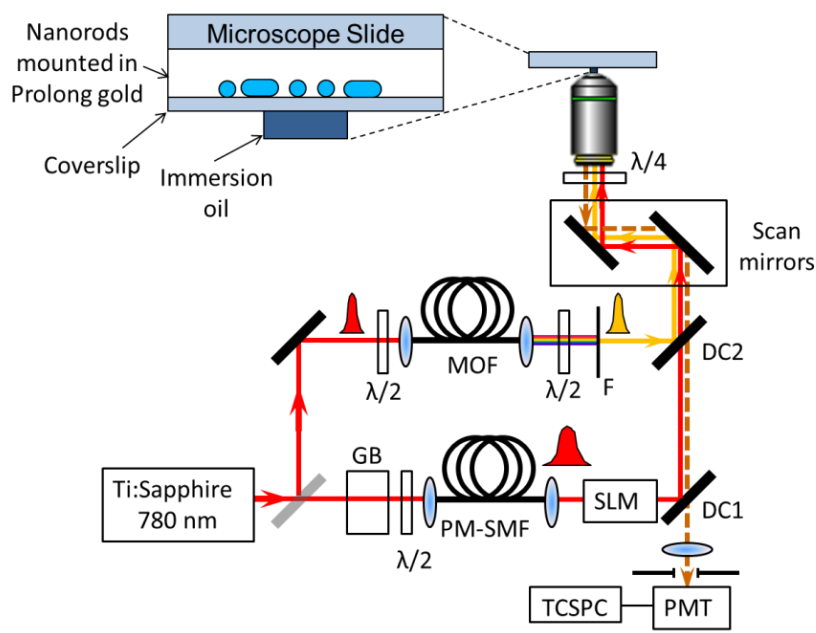

*Figure S1:* Schematic figure of the experimental setup. The nanorods are mounted in Prolong Gold on a coverslip. Components in the setup: MOF: microstructured optical fibre, F: excitation filters, λ/2: half-wave plate, λ/4: quarter-wave plate, GB: SF57 glass block, PMF-SM: polarization maintaining single mode fibre, SLM: spatial light modulator, DC1 and DC2: dichroic mirrors, PMT: photomultiplier tube, TCSPC: time correlated single photon counting unit.

Figure S1 shows a schematic of the STED-FLIM microscope. It was based on a single 80 MHz repetition rate mode-locked Ti:Sapphire laser (Spectra-Physics, Mai Tai HP) tuned to 780 nm. The beam was split into two arms by a Glan-Taylor polarizer and a half-wave plate. The excitation arm was coupled into a microstructured optical fibre (MOF) which creates a supercontinuum spanning 500 – 1000 nm, and the depletion beam was coupled into a 100 m long polarization maintaining single mode fibre (PM-SMF) via a 1 m optical path length in a SF57 glass block (GB) in order to stretch the pulse width from 100 fs to ~200 ps. The excitation wavelength was chosen to be 628/40 nm (centre wavelength/full-width at half maximum) and the collected emission was filtered by a 692/40 nm filter. The sample was imaged using an HCX PL APO 100x/1.4 oil immersion Leica objective lens. Two galvanometric scan mirrors (Yanus IV Digital Scan Head from Till Photonics, Munich, Germany) were used for scanning the beams across sample. The emitted light was coupled into a 200 µm-core fibre and the detected light was collected by a hybrid PMT (HPM-100-40 from Becker&Hickl). For FLIM, the PMT was connected to a TCSPC card (SPC-830, Becker&Hickl). The FLIM data was analysed by using the *FLIMfit* software tool developed at Imperial College London.[1]

## AuNRs size distribution

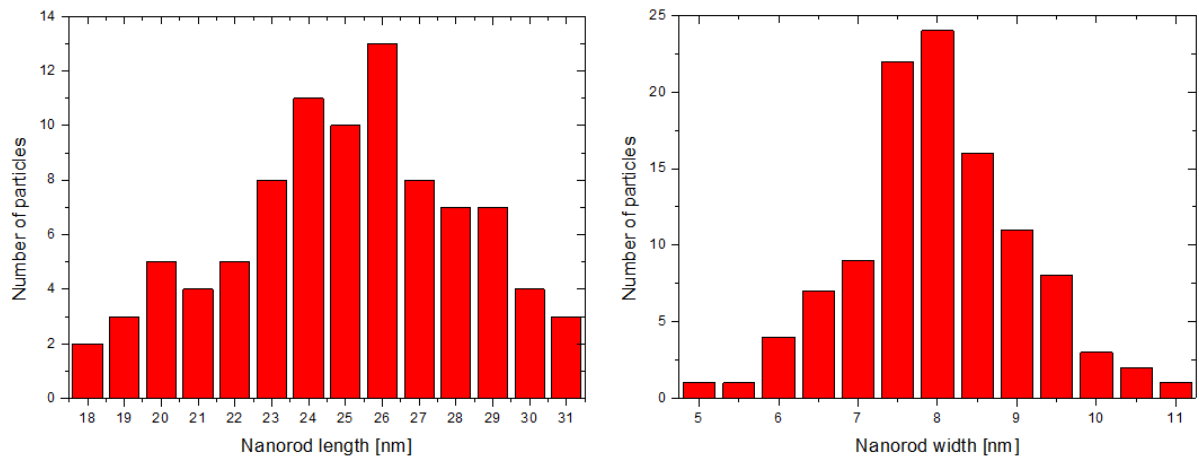

***Figure S2:*** *AuNRs size distribution determined by measurements from TEM images. Over 100 particles were measured (length and width). The average size was determined as: 26.1 ± 5.0 nm length and 8.0 ± 1.1 nm width.*

## LSPR dependence on AuNRs aspect ratio and embedding medium

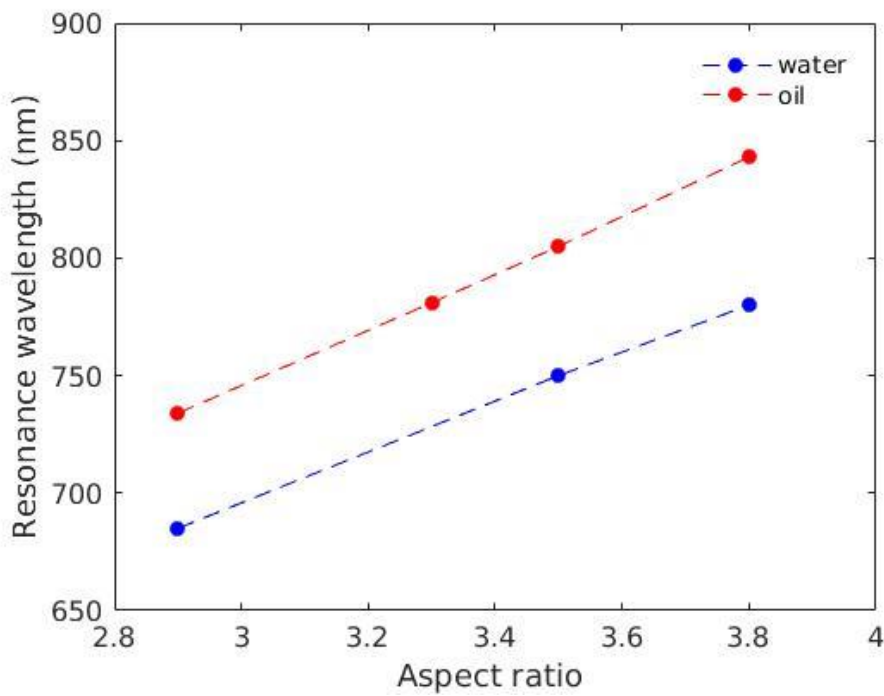

*Figure S3:* *Dependence of AuNRs longitudinal LSPR with their aspect ratio (length/width) embedded in oil (n=1.47) and water (n=1.33). The dots correspond to simulations results and the dashed lines are guides to the eye.*

Figure S3 presents the dependence of the resonance wavelength of AuNRs with their aspect ratio (length/width) for two different embedding media. When the AuNRs are immersed in oil, increasing their aspect ratio from 2.9 to 3.8 shifts the longitudinal LSPR from 740 nm to 850 nm. On the other hand, the lower refractive index of water blue-shifts the LSPR resonance.

# Lifetime measurements

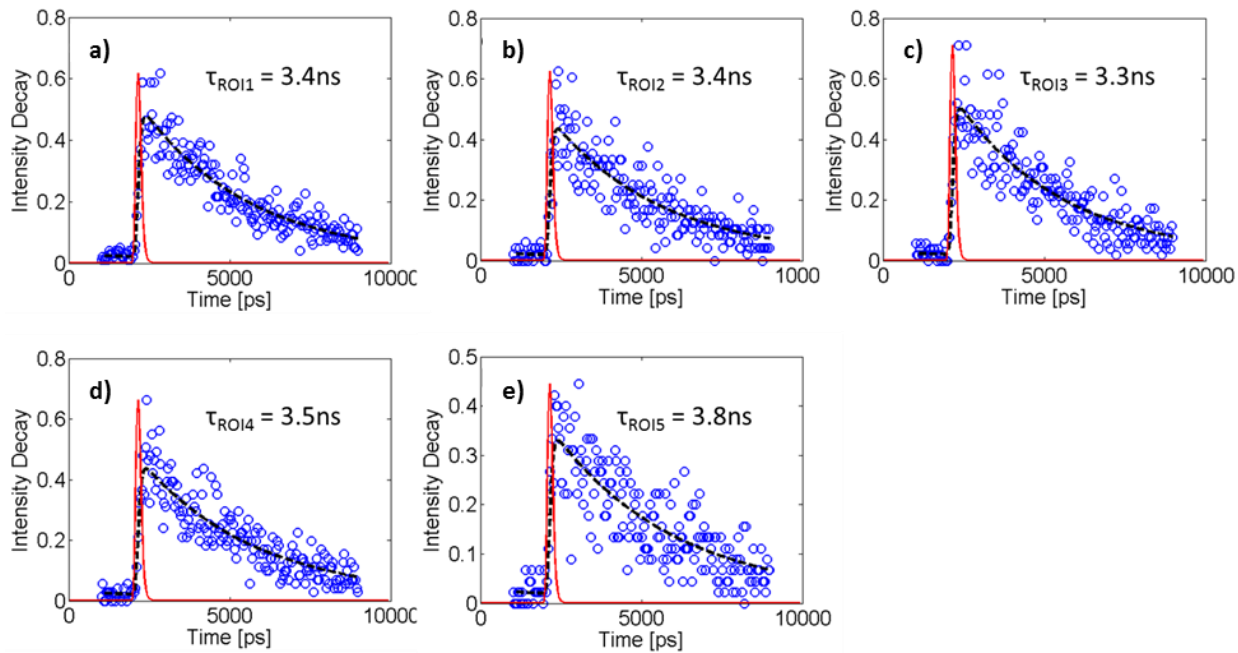

***Figure S4:*** *Examples of confocal FLIM data of nanorods and STAR 635P. The excitation wavelength was set to 628/40 nm and emission wavelength was filtered with a 692/40 nm filter. The excitation power was 5 µW, measured before the tube lens. a – e) decay data for different AuNRs. The blue circles correspond to the measured data, the dashed black lines show the decay fit from FLIM fit and the red peaks show the IRF (measured in reflection mode). Lifetimes for the areas are given as τ in the plots.*

A mean value for the lifetime was found to be 3.5 ± 0.2 ns. According to the manufacturer (http://www.abberior.com), the lifetime of STAR 635P in PBS is 3.3 ns.

# Nanoshells vs Nanorods

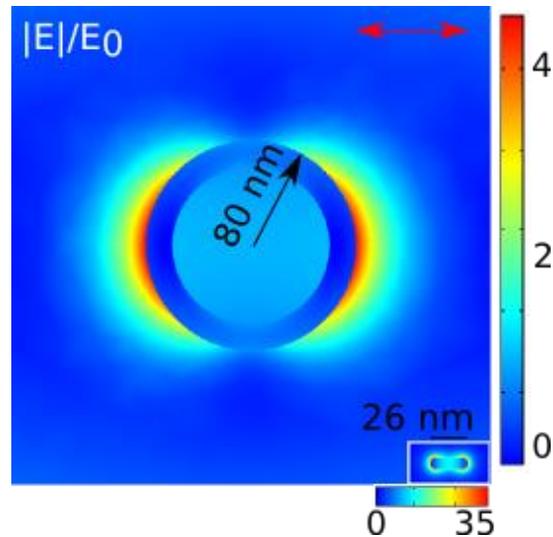

***Figure S5:*** *Computed electric field enhancement at 780 nm for silica@Au nanoparticles [2] versus AuNRs (this work). Dimensions are preserved in order to notice the reduced dimension of AuNR compared to the core@shell particle.*

Figure S5 presents a comparison of the electric field enhancement at 780 nm for the core-shell NPs used in [2] and the AuNRs used in this work. For the core@shell particles, the fluorescent dyes were incorporated inside the silica core, thus, the fluorophores experienced a ~2 fold higher field due the plasmonic Au shell. In the case of the AuNRs, as shown in Figure 2b (main text), that value increases up to 7 (for fluorophores situated 5 nm far from the surface).

## Additional NP-STED images

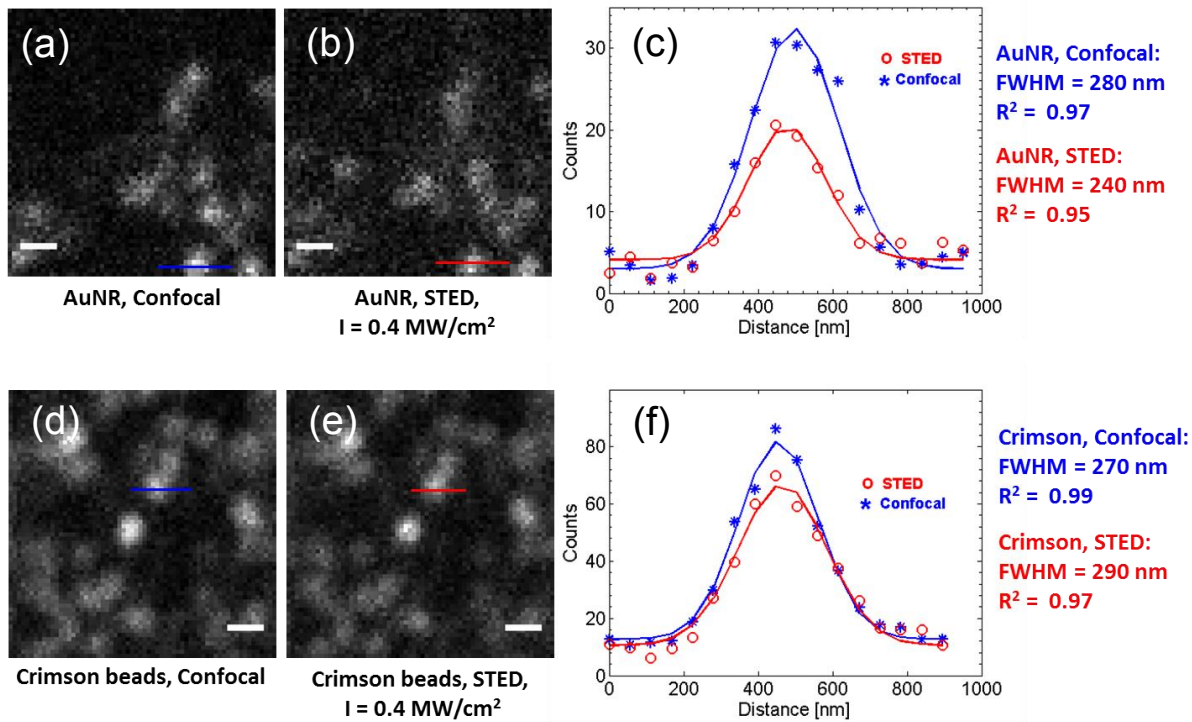

**Figure S6:** *Images a) and b) show confocal and STED images of AuNR and STAR 635P, and images d) and e) show confocal and STED images of Crimson beads. The intensity profiles are shown as blue and red lines in the images and the corresponding profile plots and Gaussian fits in c) and f). The analysed AuNRs and beads were chosen to minimise the risk of picking clusters, or areas with high fluorophore concentration. Scale bars in images correspond to 500 nm.*

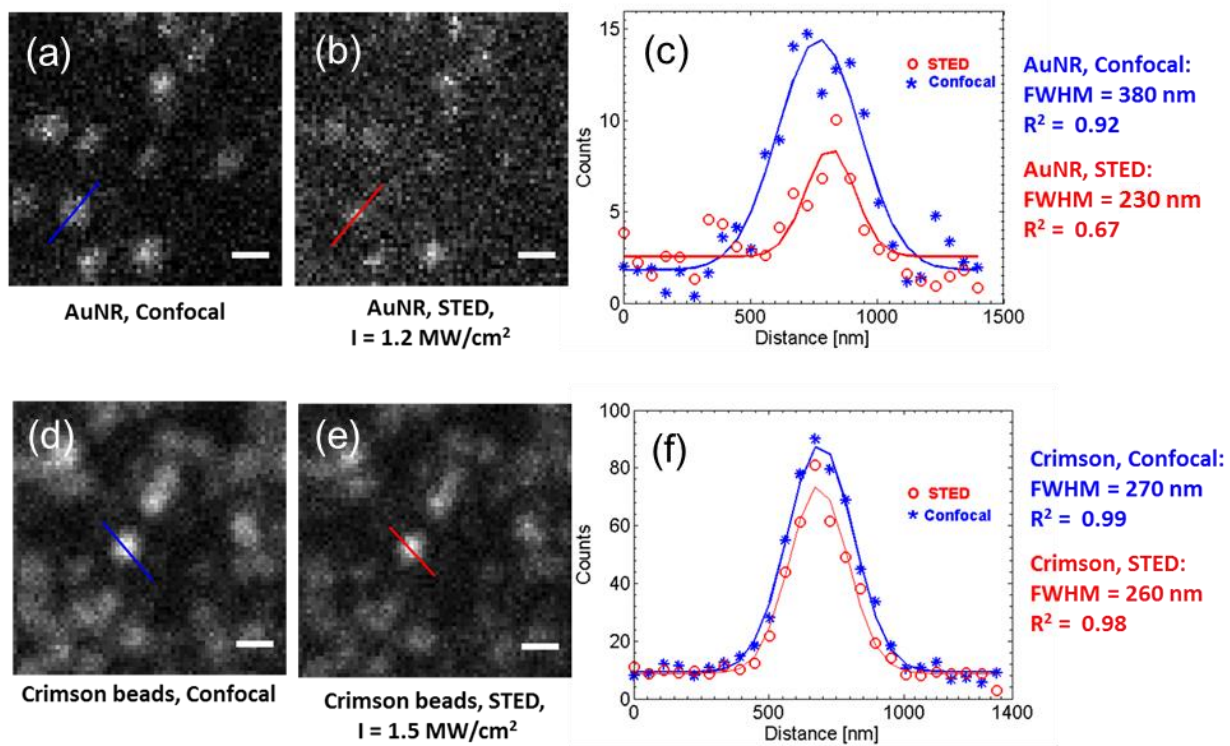

**Figure S7:** *Images a) and b) show confocal and STED images of AuNR and STAR635P, and images d) and e) show confocal and STED images of Crimson beads. The intensity profiles are shown as blue and red lines in the images and the corresponding profile plots and Gaussian fits in c) and f). The analysed AuNRs and beads were chosen to minimise the risk of picking clusters, or areas with high fluorophore concentration. Scale bars in images correspond to 500 nm.*

# Photothermal effects

As discussed in the main text, the large electromagnetic field enhancement provided by the LSPR in nanoparticles is accompanied by absorption in the metal, which in turn, results in heating of the nanoparticle and its surroundings. In the case of the AuNRs, absorption in fact dominates the extinction cross section at the LSPR. Here, we attempt to characterize the temperature increase due to the LSPR under the depletion beam using a simple modelling. In the calculations, a AuNR is illuminated with a plane wave with the same intensity as the peak intensity of the depletion beam and for the same duration a repetition rate of the STED pulse. Thus, we are disregarding the shape of the beam, and the effects of the sampling. In addition, we also note that the temperature rise calculated here should be viewed as an upper limit only. If the thermal response of the metal, which generically results in an increase of the imaginary part of the permittivity, is taken into account, the obtained temperatures can be substantially lower [4].

Figure S8(a) shows the temperature rise calculated for a single AuNR under pulsed illumination (see below for calculation details) for conditions corresponding to those in the STED microcopy experiments. As shown in the plot, the AuNR temperature rises quickly up to a maximum value during the duration of the 200 ps STED pulse (specifically, for $I_0$ = 1 MW/cm² the temperature increases by 200 degrees). Note here that a peak intensity of 1 MW/cm² corresponds to an average power density of $I_o \, \tau_{STED} \, f$ = 0.02 MW/cm², being $\tau_{STED}$ the duration of the pulse (200 ps) and $f$ the repetition rate (80 MHz). On the other hand, the low duty cycle of the STED pulse means that the system has time to dissipate heat: the temperature goes back to ambient in ~6 ns (about half the pulse period of 12 ns), which is longer than the time scale plotted in the figure. The inset panel presents the spatial profile of the temperature in the vicinity of the AuNR at the end of the pulse (t = 0.2 ns), showing how the temperature greatly increases in the near field of the particle while it rapidly falls away from the metal surface. In Figure S8(b) we present the calculated temperature rise for longer pulses (1ns) than those used in the experiments, showing how increasing the pulse duration results in a lower peak temperature (over 50% reduction compared to the 200ps pulses). Importantly, since the pulse energy is kept, the same resolution is achieved while reducing the temperature increase.

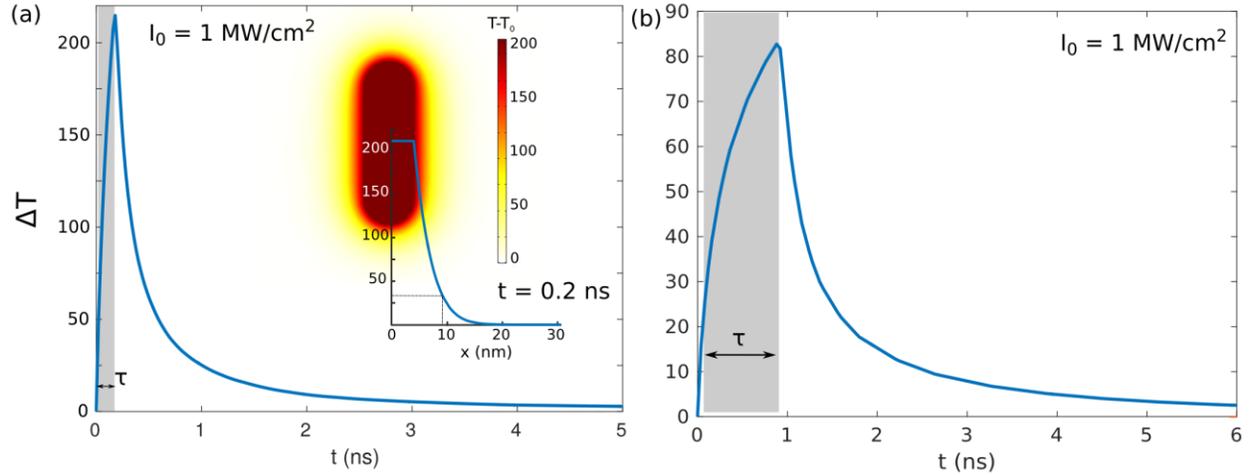

***Figure S8: (a)*** *Heat dissipation of the rods when illuminated with a circularly polarized pulsed plane wave: peak power 1MW/cm², illumination time τ = 200ps, repetition rate 80MHz (period 12 ns). The inset plot shows the temperature spatial distribution at the end of the pulse, t = τ, (colour map). The temperature decay away from the nanoparticle is also shown in a line plot.* ***(b)*** *Temperature dissipation for the case of: peak power 1MW/cm², illumination time τ = 1 ns, repetition rate 80MHz (period 12 ns).*

**Temperature calculations:** The photothermal heating of the gold nanorods under the modulated laser illumination was studied by solving the heat diffusion equation,

$$C_p(\mathbf{r})\rho(\mathbf{r})\frac{\partial T(\mathbf{r})}{\partial t} + \nabla[-\kappa(\mathbf{r}) \cdot \nabla T(\mathbf{r})] = q(\mathbf{r})$$

where T is the temperature, $C_p$ the specific heat (129 J·kg/K for gold and 2000 J·kg/K for oil), $\rho$ the density (1,93 · 10⁴ kg/m³ for gold and 8,44 · 10² kg/m³ for oil), $\kappa$ stands for thermal conductivity (317 W·m/K for gold and 0,145 W·m/K for oil). The heat source is due to the optical absorption of the particle,

$$q(\mathbf{r}) = \frac{1}{2}\varepsilon_0\omega\text{Im}(\varepsilon_{Au})|\mathbf{E}(\mathbf{r})|^2$$

which is modulated following the on-off cycle of the STED laser used in the experiments (illumination time $\tau = 200$ ps, repetition rate 80 MHz). The heat equation was solved by means of the finite element method [Comsol Multiphysics, https://www.comsol.com/comsol-multiphysics].

# AuNRs-STED anisotropy

We have considered that most of our AuNRs are sitting flat for the measurements performed on glass cover slips (Figure 3), while this is not the case for the bioimaging example (Figure 4). In order to account for the 3D orientation of the AuNRs inside the cells we have estimated the resolution improvement as a function of the AuNRs orientation.

The vectorial diffraction theory of Richards and Wolf [5], [6] was used to calculate the x, y and z polarized electric field components of a doughnut shaped depletion focus of wavelength 780nm, for an oil immersion objective lens with a numerical aperture of 1.4. The projection of the rod's long axis onto the x, y and z axes (see Figure S9) was then used to calculate the enhancement of each of the electric field components separately. The effective intensities of the three polarization components are then,

$$I_{eff-x} = |(1 + (\eta - 1)cos\theta cos\phi)E_x|^2$$

$$I_{eff-y} = |(1 + (\eta - 1)sin\theta cos\phi)E_y|^2$$

$$I_{eff-y} = |(1 + (\eta - 1)sin\phi)E_z|^2$$

where $\theta$ is the angle between the x axis and the long axis of the rod, $\phi$ is the angle between the x-y plane and the long axis of the rod, $\eta$ is the field enhancement and $E_x$, $E_y$ and $E_z$ are the electric field components of the incident (unenhanced) STED PSF. The total effective depletion intensity is then

$$I_{eff-xyz} = I_{eff-x} + I_{eff-y} + I_{eff-z}$$

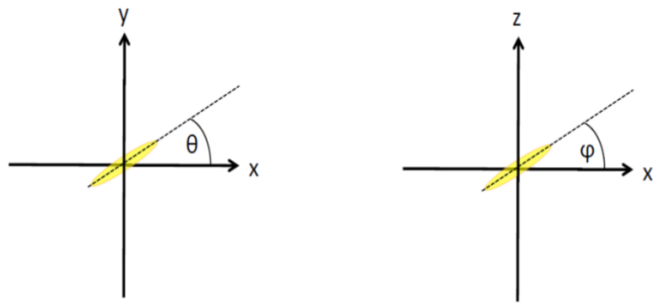

*Figure S9: Projection of the AuNRs long axis onto each of the x, y and z axes used to calculate the effective intensities of the STED and emission PSFs. The in-plane variation is characterized by the angle $\theta$ and the out-of-plane one by $\phi$.*

Figure S10 shows the variation of the depletion PSF intensity for an in-plane rod ($\phi = 0$) with increasing angle $\theta$. Also shown are the corresponding emission PSFs for a peak incident (unenhanced) STED intensity of 5 times the saturation intensity of the fluorophore [7] and the FWHMs of Gaussian fits to line profiles through the centres of the emission PSFs

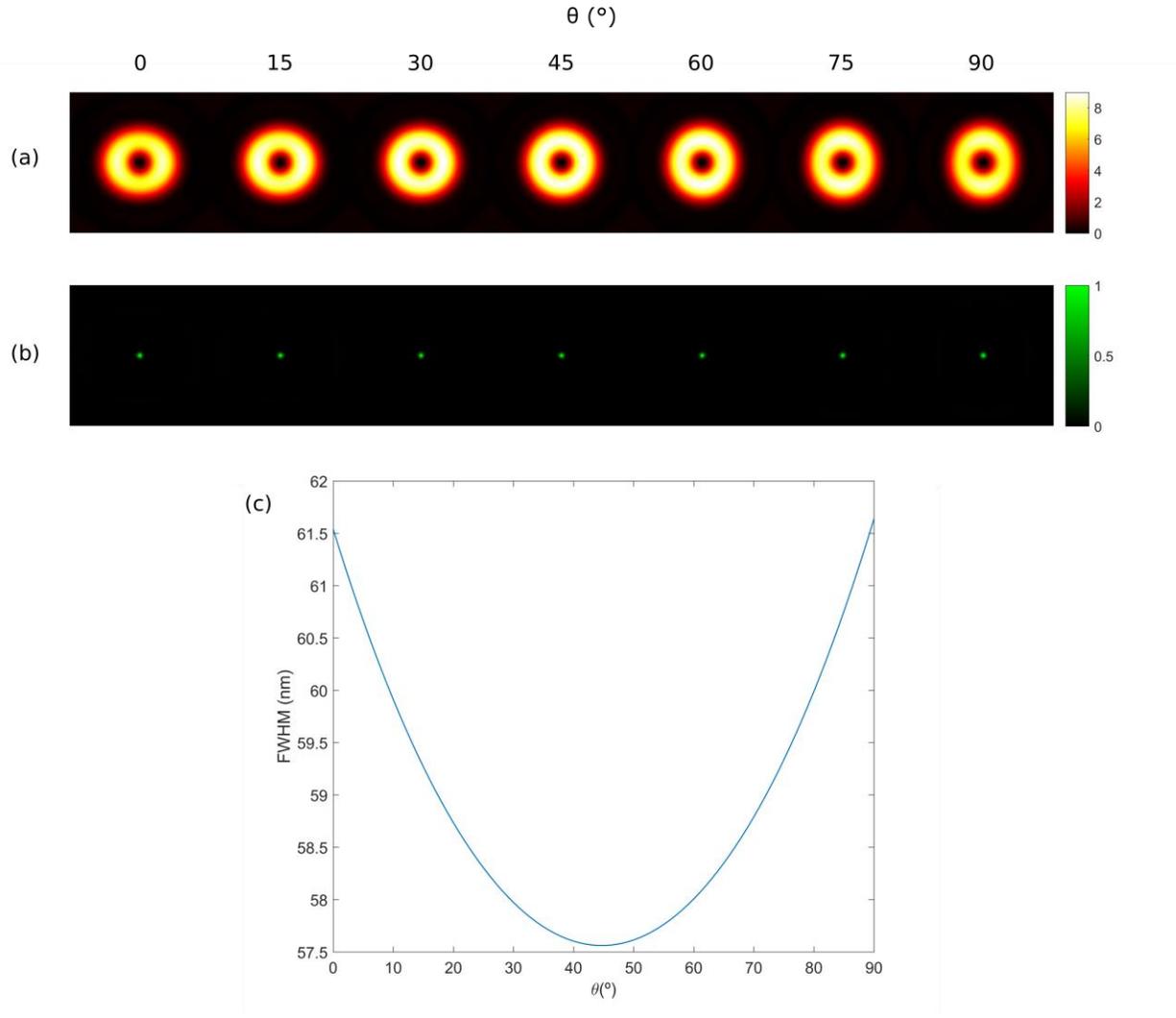

*Figure S10: a) Effective STED PSFs, b) Effective emission PSFs and c) emission PSF FWHM with varying rod angle $\phi$. Calculated for NA = 1.4, depletion wavelength = 780nm, excitation wavelength = 635nm, peak incident STED intensity = $5I_{sat}$ and field enhancement $\eta = 4$.*

Figure S11 shows the variation of the depletion PSF intensity for varying angle $\phi$. Again, the corresponding emission PSFs for a peak incident (unenhanced) STED intensity of 5 times the saturation intensity of the fluorophore and the FWHMs of Gaussian fits to line profiles through the centres of the emission PSFs are also shown.

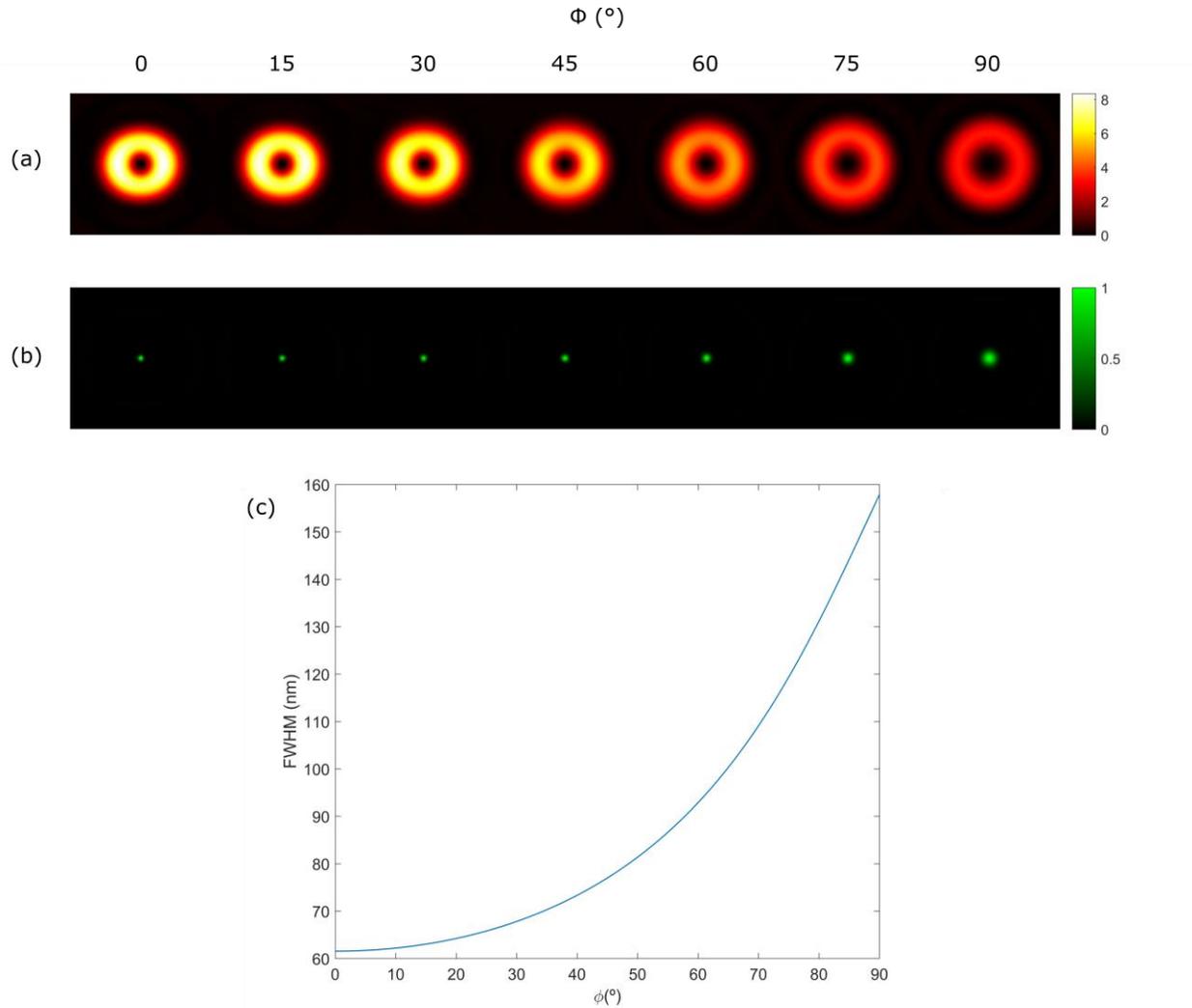

***Figure S11:*** *a) Effective STED PSFs, b) Effective emission PSFs and c) emission PSF FWHM with varying rod angle $\phi$. Calculated for NA = 1.4, depletion wavelength = 780nm, excitation wavelength = 635nm, peak incident STED intensity = $5I_{sat}$ and field enhancement $\eta = 4$.*

As expected, variations are larger with the out-of-plane angle ($\phi$) than with the in-plane one ($\theta$) as we are using circularly polarized light. In Figures S10 (b) and (c) it can be seen that there is a slight variation of the effective emission PSF as the in-plane rod angle $\theta$ increases, with the PSF FHWM showing a maximum deviation of approximately 7% from the minimum value. This is not the case as the orientation of the rod is moved out of the x-y plane. As $\phi$ increases there is a much more profound effect on the emission PSF FWHM. If the long axis of the rod is aligned along the z-axis (i.e. $\phi = 90$) of the experiment then, for the parameters used for these calculations, the resultant emission PSF has a FWHM approximately 160% larger than is seen for an in-plane rod ($\phi = 90$).

## Supplementary References

1. Warren, S.C., et al., *Rapid Global Fitting of Large Fluorescence Lifetime Imaging Microscopy Datasets.* PLoS ONE, 2013. **8**(8): p. e70687.

2. Sonnefraud, Y., et al., *Experimental Proof of Concept of Nanoparticle-Assisted STED.* Nano Letters, 2014. **14**(8): p. 4449-4453.

3. Shen, P.-T., et al., *Temperature- and -roughness dependent permittivity of annealed/unannealed gold films.* Optics Express, 2016. **24**(17): p. 19254-19263.

4. Yonatan Sivan, S.-W.C., *Nonlinear plasmonics at high temperatures.* Nanophotonics, 2016. **Accepted**.

5. B. Richards and E. Wolf, "Electromagnetic Diffraction in Optical Systems. II. Structure of the Image Field in an Aplanatic System," *Proc. R. Soc. A Math. Phys. Eng. Sci.*, vol. 253, no. 1274, pp. 358–379, Dec. 1959.

6. B. R. Boruah and M. A. A. Neil, "Focal field computation of an arbitrarily polarized beam using fast Fourier transforms," *Opt. Commun.*, vol. 282, no. 24, pp. 4660–4667, Dec. 2009.

7. M. Leutenegger, C. Eggeling, and S. W. Hell, "Analytical description of STED microscopy performance.," *Opt. Express*, vol. 18, no. 25, pp. 26417–29, Dec. 2010.

## References:

[1] Hell, S.W. and J. Wichmann, Breaking the diffraction resolution limit by stimulated emission: stimulated-emission-depletion fluorescence microscopy. Optics Letters, 1994. 19(11): p. 780-782.

[2] T. A. Klar, S. Jakobs, M. Dyba, A. Egner, and S. W. Hell, "Fluorescence microscopy with diffraction resolution barrier broken by stimulated emission," Proc. Natl. Acad. Sci. U.S.A. 2000. 97(15), 8206–8210.

[3] S. W. Hell, "Toward fluorescence nanoscopy," Nat. Biotechnol. 2003. 21(11), 1347–1355.

[4] M. Hofmann, C. Eggeling, S. Jakobs, and S. W. Hell, "Breaking the diffraction barrier in fluorescence microscopy at low light intensities by using reversibly photoswitchable proteins," Proc. Natl. Acad. Sci. U.S.A. 2005. 102(49), 17565–17569.

[5] E. Betzig, G. H. Patterson, R. Sougrat, O. W. Lindwasser, S. Olenych, J. S. Bonifacino, M. W. Davidson, J. Lippincott-Schwartz, and H. F. Hess, Science 2006. 313, 1642–1645.

[6] S. T. Hess, T. P. K. Girirajan, and M. D. Mason, Biophys, J. 2006. 91, 4258–4272.

[7] M. J. Rust, M. Bates, and X. Zhuang, Nat. Methods 2006. 3, 793–795.

[8] Orrit, M., Nobel Prize in Chemistry: Celebrating optical nanoscopy. Nat Photon, 2014. 8(12): p. 887-888.

[9] Willig, K.I., et al., Nanoscale resolution in GFP-based microscopy. Nat Meth, 2006. 3(9): p. 721-723.

[10] S. Berning, K. I. Willig, H. Steffens, P. Dibaj, and S. W. Hell, "Nanoscopy in a Living Mouse Brain," Science 2012. 335(6068), 551.

[11] E. Rittweger, K. Y. Han, S. E. Irvine, C. Eggeling, and S. W. Hell, "STED microscopy reveals crystal colour centres with nanometric resolution," Nat. Photonics 2009. 3(3), 144–147.

[12] Bingen, P., et al., Parallelized STED fluorescence nanoscopy. Optics Express, 2011. 19(24): p. 23716-23726.

[13] B. Yang, F. Przybilla, M. Mestre, J.-B. Trebbia, and B. Lounis, "Large parallelization of STED nanoscopy using optical lattices," Opt. Express 2014. 22(5), 5581–5589.

[14] Hanne, J., et al., STED nanoscopy with fluorescent quantum dots. Nat Commun, 2015. 6

[15] Sivan, Y., et al., Nanoparticle-Assisted Stimulated-Emission-Depletion Nanoscopy. ACS Nano, 2012. 6(6): p. 5291-5296.

[16] Sivan, Y., Performance improvement in nanoparticle-assisted stimulated-emission-depletion nanoscopy. Applied Physics Letters, 2012. 101(2): p. 021111.

[17] Sonnefraud, Y., et al., Experimental Proof of Concept of Nanoparticle-Assisted STED. Nano Letters, 2014. 14(8): p. 4449-4453.

[18] Giannini, V., et al., Plasmonic Nanoantennas: Fundamentals and Their Use in Controlling the Radiative Properties of Nanoemitters. Chemical Reviews, 2011. 111(6): p. 3888-3912.

[19] Halas, N.J., et al., Plasmons in Strongly Coupled Metallic Nanostructures. Chemical Reviews, 2011. 111(6): p. 3913-3961.

[20] Wu, H-Y., Ultrasmall all-optical plasmonic switch and its application to superresolution imaging. Scientific Reports, 2016. 6, 24293.

[21] Chu, S-W., Measurement of a Saturated Emission of Optical Radiation from Gold Nanoparticles: Application to an Ultrahigh Resolution Microscope. Physical Review Letters, 2014. 112, 017402.

[22] Lee, S. A., Ponjavic, A., Siv, C., Lee, S. F. and Biteen J. S., Nanoscopic Cellular Imaging: Confinement Broadens Understanding. ACS Nano, 2016. Just accepted, 10.1021/acsnano.6b02863.

[23] Blythe, K.L., Titus, E.J., and Willets K.A., The effects of tuning fluorophore density, identity, and spacing on reconstructed images in super-resolution imaging of fluorophore-labeled gold nanorods. Journal of Physical Chemistry C, 2015. 119, 28099.

[24] Albanese, A., P.S. Tang, and W.C.W. Chan, The Effect of Nanoparticle Size, Shape, and Surface Chemistry on Biological Systems. Annual Review of Biomedical Engineering, 2012. 14(1): p. 1-16.

[25] Ni, M., et al., Fluorescent probes for nanoscopy: four categories and multiple possibilities. Journal of Biophotonics, 2016. p:1-13

[26] Maier, S.A., Plasmonics: Fundamentals and Applications. 1 ed. 2007: Springer US. 224.

[27] Eustis, S. and M.A. El-Sayed, Why gold nanoparticles are more precious than pretty gold: Noble metal surface plasmon resonance and its enhancement of the radiative and nonradiative properties of nanocrystals of different shapes. Chemical Society Reviews, 2006. 35(3): p. 209-217.

[28] Link, S., M.B. Mohamed, and M.A. El-Sayed, Simulation of the Optical Absorption Spectra of Gold Nanorods as a Function of Their Aspect Ratio and the Effect of the Medium Dielectric Constant. The Journal of Physical Chemistry B, 1999. 103(16): p. 3073-3077.

[29] Balzarotti, F. and F.D. Stefani, Plasmonics Meets Far-Field Optical Nanoscopy. ACS Nano, 2012. 6(6): p. 4580-4584.

[30] Anger, P., P. Bharadwaj, and L. Novotny, Enhancement and Quenching of Single-Molecule Fluorescence. Physical Review Letters, 2006. 96(11): p. 113002.

[31] Darst, S.A., et al., Two-dimensional crystals of streptavidin on biotinylated lipid layers and their interactions with biotinylated macromolecules. Biophysical Journal, 1991. 59(2): p. 387-396.

[32] Leutenegger, M. et al., Analytical description of STED microscopy performance,. Optics Express, 2010. 18(25): p. 26417-26429.

[33] Fixler, D., T. Nayhoz, and K. Ray, Diffusion Reflection and Fluorescence Lifetime Imaging Microscopy Study of Fluorophore-Conjugated Gold Nanoparticles or Nanorods in Solid Phantoms. ACS Photonics, 2014. 1(9): p. 900-905.

[34] Foreman, M.R., et al., Independence of plasmonic near-field enhancements to illumination beam profile. Physical Review B, 2012. 86(15): p. 155441.

[35] Wolfbeis, O.S., An overview of nanoparticles commonly used in fluorescent bioimaging. Chemical Society Reviews, 2015. 44(14): p. 4743-4768.

[36] Pustovalov, V.K., Light-to-heat conversion and heating of single nanoparticles, their assemblies, and the surrounding medium under laser pulses. RSC Advances, 2016. 6(84): p. 81266-81289.

[37] Bordenave M. D., Balzarotti, F., Stefani, F.D., and Hell, S.W., STED nanoscopy with wavelengths at the emission maximum Journal of Physics D: Applied Physics, 2016. 49, 36.

[38] G. Vicidomini, G. Moneron, C. Eggeling, E. Rittweger and S. W. Hell, STED with wavelengths closer to the emission máximum. Optics Express, 2012. 20(5): p 4154-4162.

[39] 39 Zhou, N., et al., Silver Nanoshell Plasmonically Controlled Emission of Semiconductor Quantum Dots in the Strong Coupling Regime. ACS Nano, 2016. 10(4): p. 4154-4163.

[40] Rakovich, A., P. Albella, and S.A. Maier, Plasmonic Control of Radiative Properties of Semiconductor Quantum Dots Coupled to Plasmonic Ring Cavities. ACS Nano, 2015. 9(3): p. 2648-2658.

[41] Han, K. Y. and Ha, T. Dual-color three-dimensional STED microscopy with a single high-repetition-rate laser. Optics Letters, 2015. 40(11): p.2653-2656.

[42] Johnson, P.B. and R.W. Christy, Optical Constants of the Noble Metals. Physical Review B, 1972. 6(12): p. 4370-4379.

[43] Auksorius, E., et al., Stimulated emission depletion microscopy with a supercontinuum source and fluorescence lifetime imaging. Optics Letters, 2008. 33(2): p. 113-115.

[44] Lenz, M.O., et al., 3-D stimulated emission depletion microscopy with programmable aberration correction. Journal of Biophotonics, 2014. 7(1-2): p. 29-36.

[45] Vicidomini, G., et al., STED Nanoscopy with Time-Gated Detection: Theoretical and Experimental Aspects. Plos One, 2013. 8(1).

[46] Warren, S.C., et al., Rapid Global Fitting of Large Fluorescence Lifetime Imaging Microscopy Datasets. PLoS ONE, 2013. 8(8): p. e70687.

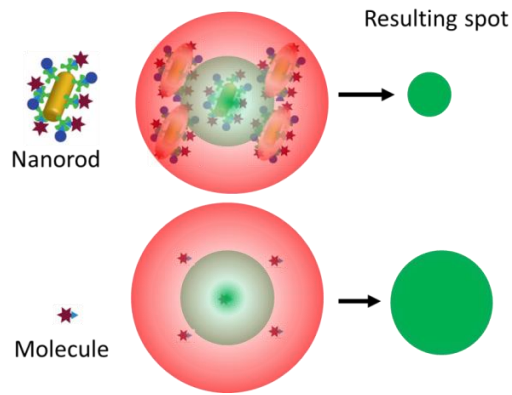